\documentclass[conference]{IEEEtran} 
\IEEEoverridecommandlockouts
\usepackage[binary-units,per-mode=symbol,per-symbol=p]{siunitx}
\usepackage{cite}
\usepackage{adjustbox}
\usepackage[ruled,vlined]{algorithm2e}
\usepackage{amsmath,amssymb,amsfonts}
\usepackage{algorithmic}
\usepackage{graphicx}
\usepackage{textcomp}
\usepackage{multirow}
\usepackage[colorlinks=false, pdfborder={0 0 0}, breaklinks]{hyperref}
\usepackage{xcolor}
\usepackage{balance}
\usepackage{enumitem}
\usepackage{setspace}
\usepackage{wrapfig}
\usepackage{listings}
\usepackage{color}
\usepackage{caption}
\usepackage{subcaption}
\usepackage{todonotes}

\def\BibTeX{{\rm B\kern-.05em{\sc i\kern-.025em b}\kern-.08em
    T\kern-.1667em\lower.7ex\hbox{E}\kern-.125emX}}

\usepackage{amsmath}

\usepackage{ctable}

\newcommand{\etal}{\emph{et al.}\xspace}

\newcommand{\eg}{\emph{e.g.}, }
\newcommand{\etc}{\emph{etc. }}

\newcommand{\BDRP}{$BDR_{\text{P}}$\xspace}
\newcommand{\BDRX}{$BDR_{\text{X}}$\xspace}
\newcommand{\delTP}{$\Delta T_{\text{P}}$\xspace}
\newcommand{\delTS}{$\Delta T_{\text{S}}$\xspace}

\begin{document}

\title{Block-Partitioning Strategies for Accelerated Multi-rate Encoding in Adaptive VVC Streaming}

\author{\IEEEauthorblockN{Vignesh V Menon\IEEEauthorrefmark{1}, Adam Wieckowski\IEEEauthorrefmark{1}, Yiquin Liu\IEEEauthorrefmark{1}, Benjamin Bross\IEEEauthorrefmark{1}, Detlev Marpe\IEEEauthorrefmark{1}
}
\IEEEauthorblockA{\IEEEauthorrefmark{1}Video Communication and Applications department, Fraunhofer HHI, Berlin, Germany}
}

\maketitle

\begin{abstract}
The demand for efficient multi-rate encoding techniques has surged with the increasing prevalence of ultra-high-definition (UHD) video content, particularly in adaptive streaming scenarios where a single video must be encoded at multiple bitrates to accommodate diverse network conditions. While Versatile Video Coding (VVC) significantly improves compression efficiency, it introduces considerable computational complexity, making multi-rate encoding a resource-intensive task. This paper examines coding unit (CU) partitioning strategies to minimize redundant computations in VVC while preserving high video quality. We propose single- and double-bound approaches, leveraging CU depth constraints from reference encodes to guide dependent encodes across multiple QPs. These methods are evaluated using VVenC with various presets, demonstrating consistent improvements in encoding efficiency. Our methods achieve up to \SI{11.69}{\percent} reduction in encoding time with minimal bitrate overhead ($<$\SI{0.6}{\percent}). Comparative Pareto-front (PF) analysis highlights the superior performance of multi-rate approaches over existing configurations. These findings validate the potential of CU-guided strategies for scalable multi-rate encoding in adaptive streaming.
\end{abstract}


\section{Introduction}
The proliferation of high-definition (HD) and ultra-high-definition (UHD) video content has dramatically increased the demand for efficient video encoding techniques. With the advent of streaming services and the widespread availability of diverse network conditions, it has become essential to develop encoding methods that efficiently compress video data and dynamically adapt to varying bandwidth and device capabilities. Versatile Video Coding (VVC)~\cite{vvc_ref}, introduced by the Joint Video Experts Team (JVET), represents the latest advancement in video compression standards, offering significant improvements in compression efficiency compared to its predecessor, High Efficiency Video Coding (HEVC)~\cite{HEVC}. However, these improvements come with a substantial increase in computational complexity, which poses a significant challenge for real-time and adaptive streaming applications.

\begin{figure}[t]
    \centering
\begin{subfigure}{0.475\linewidth}
    \centering
    \includegraphics[width=\textwidth]{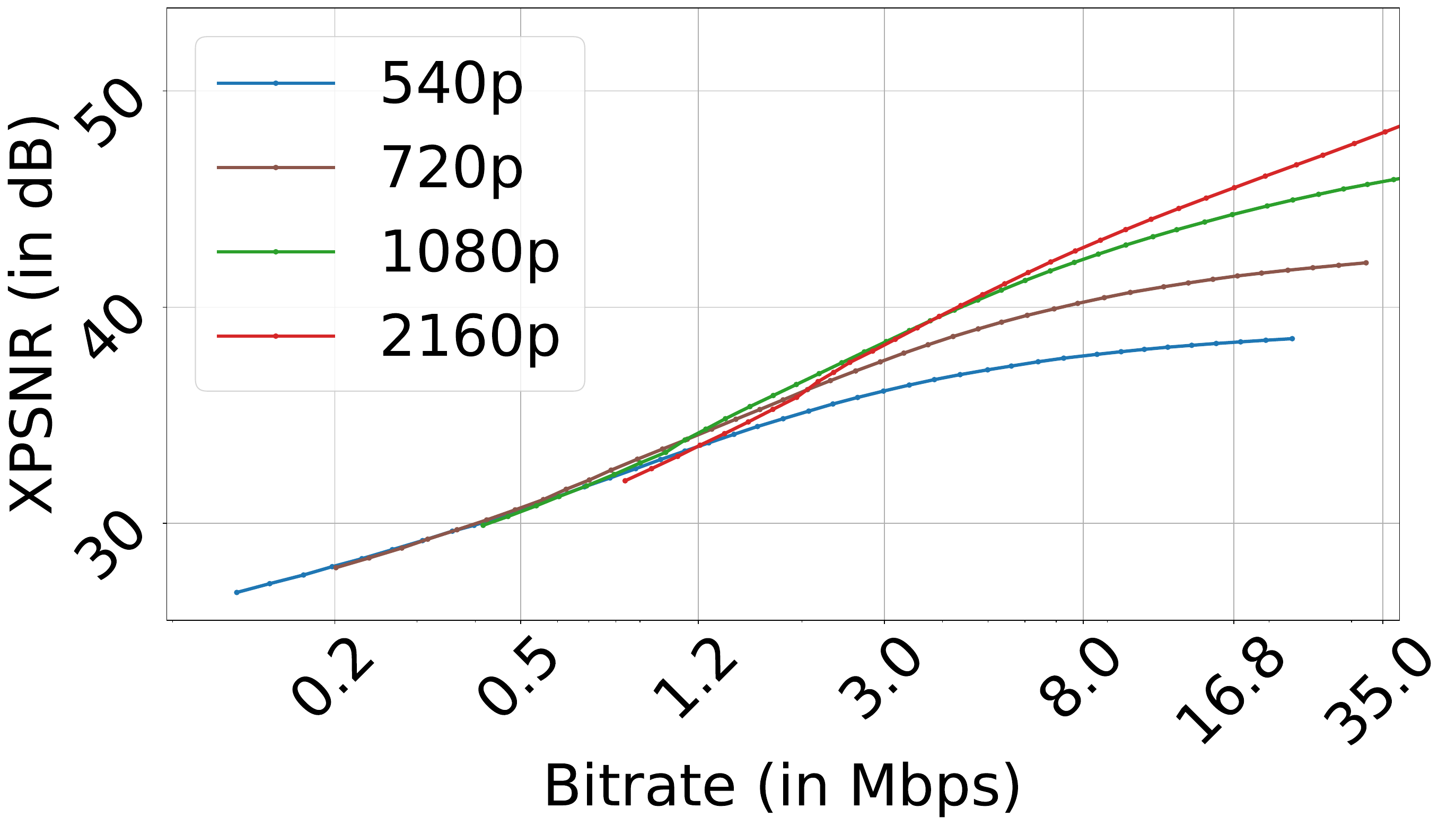}
    \includegraphics[width=\textwidth]{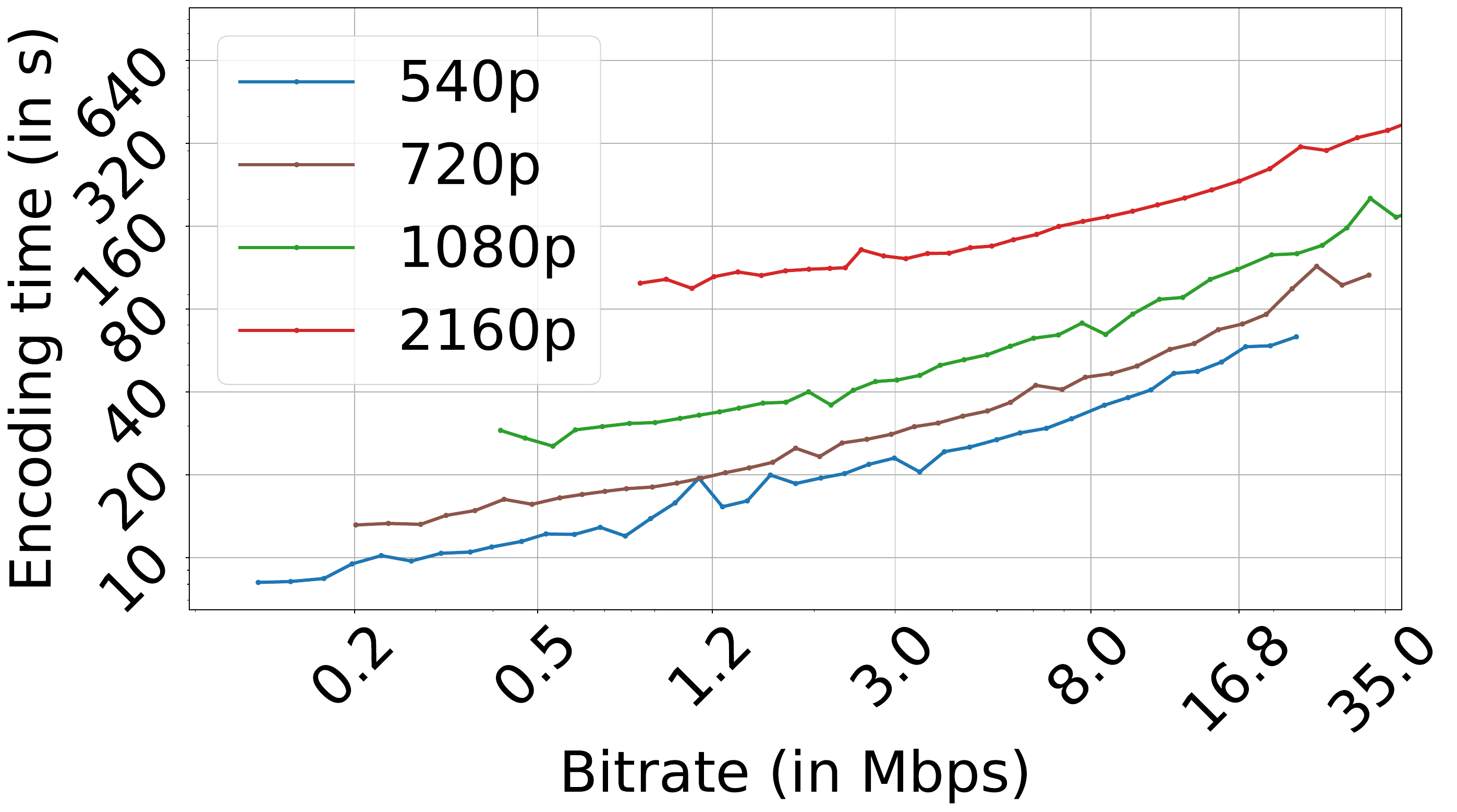}    
    \caption{\texttt{Sequence 0265}}
\end{subfigure}
\begin{subfigure}{0.475\linewidth}
    \centering
    \includegraphics[width=\textwidth]{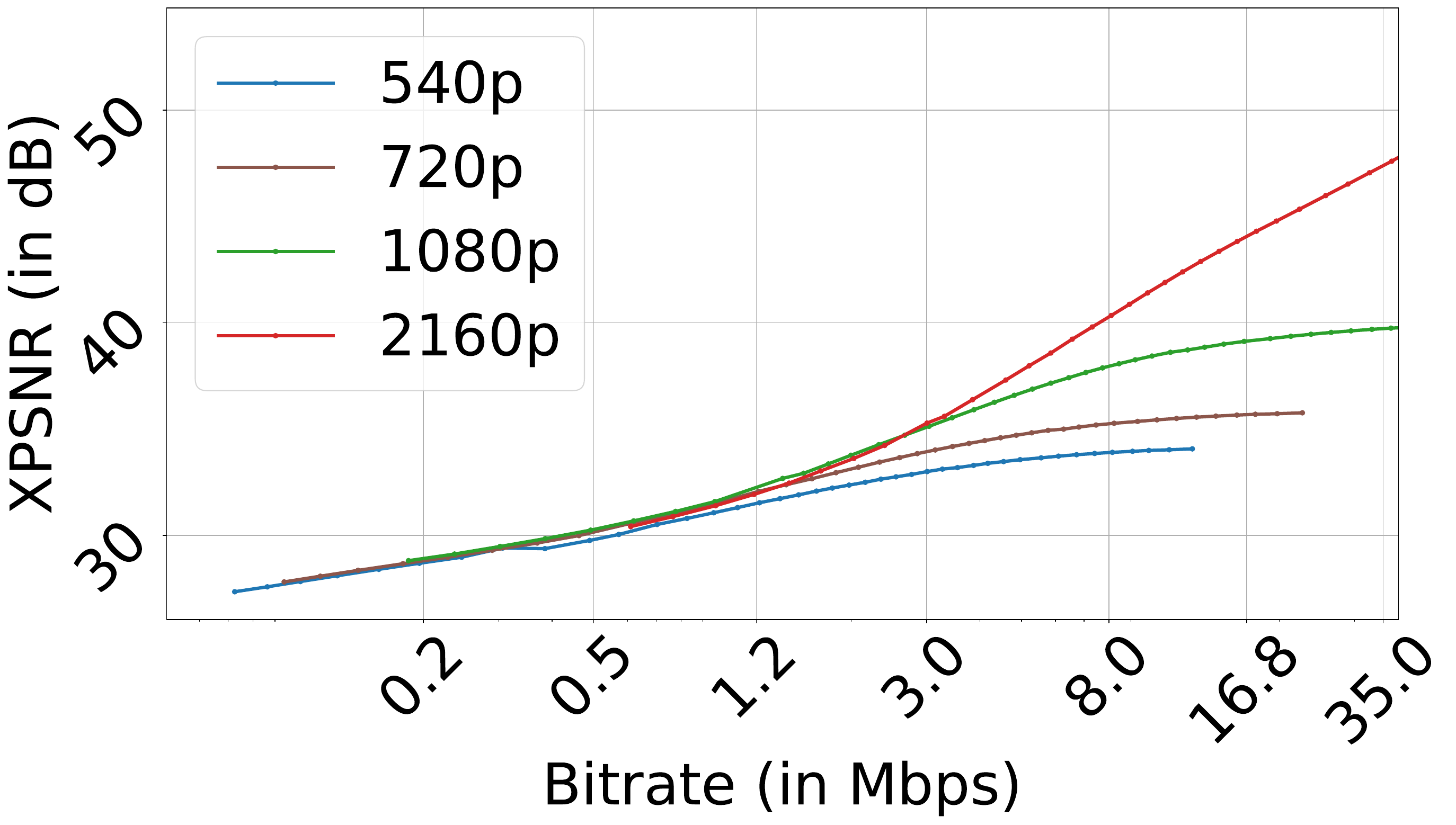}
    \includegraphics[width=\textwidth]{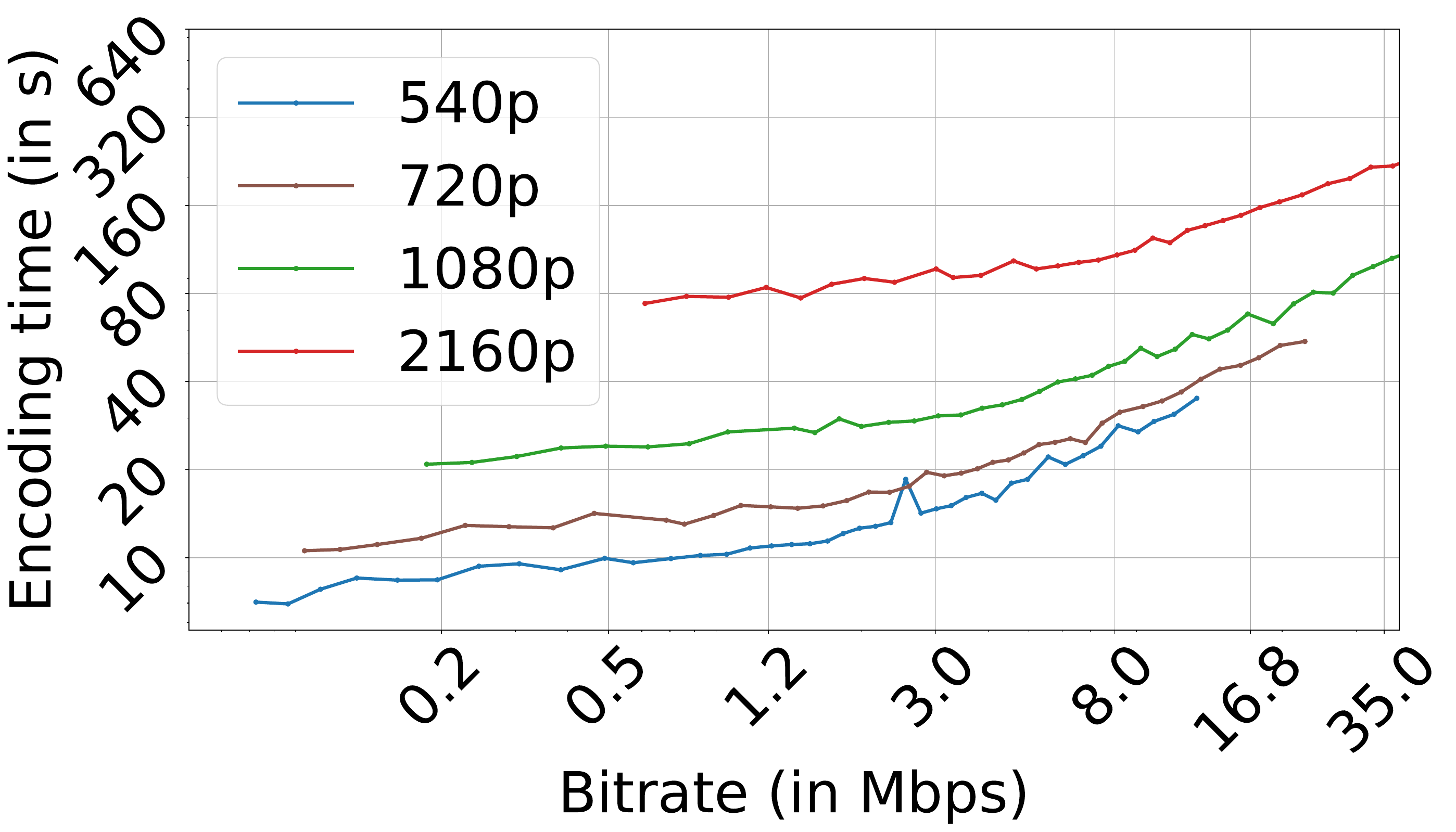}    
    \caption{\texttt{Sequence 0276}}
\end{subfigure}
\caption{Rate-Quality and Rate-Encoding time curves for VVenC~\cite{vvenc_ref} across spatial resolutions for selected sequences from the Inter-4K dataset~\cite{inter4k_ref}.}
\vspace{-1.99em}
\label{fig:intro_complexity}
\end{figure}

In adaptive streaming, a video is encoded at multiple bitrates to ensure a seamless viewing experience regardless of the user’s network conditions~\cite{mpeg_dash_ref}. In cloud-based encoding pipelines, such as those deployed by commercial streaming providers (\eg Netflix, YouTube), videos are often encoded across 10–15 ladder points. This multi-rate encoding strategy enables the streaming server to switch between different quality levels based on the current bandwidth and the network condition of the user, thereby preventing buffering and ensuring a consistent quality of experience~\cite{bentaleb_survey_2019}. Recently, Netflix introduced per-title encoding approaches to optimize rate-quality curves in the video-on-demand streaming, where a video (referred to as a \emph{title}) is encoded $\tilde{r}\times \tilde{q}$ times to determine the convex-hull~\cite{netflix_paper,menon_jnd-aware_2024}, as shown in Fig.~\ref{fig:intro_complexity}, where $\tilde{r}$ and $\tilde{q}$ represent the number of supported resolutions and quantization parameters (QP), respectively. This further enhanced the computational burden of encoding servers, where a single video is encoded multiple times at different bitrates and resolutions, particularly with the sophisticated encoding techniques employed in VVC. Thus, even a \SI{10}{\percent} reduction in encoding time per video can translate into substantial cost and energy savings~\cite{ramasubbu_modeling_2022}. 

Encoding a video at multiple bitrates typically involves independently processing it at each target bitrate, resulting in redundant computations, particularly during the CU partitioning stage~\cite{Schroeder_ref, schroeder_efficient_2018}. This redundancy increases encoding time and demands more processing power. To address these challenges, there is growing interest in techniques that reuse information from one encoding pass to accelerate subsequent encodings at different bitrates, thereby reducing overall computational load. Schroeder~\etal~\cite{Schroeder_ref} examined block structure similarities across resolutions in HEVC encoding and proposed deriving low-resolution block structures from high-resolution encodings. Cai~\etal~\cite{hevc_multirate_x265} proposed a multi-rate encoding system leveraging depth correlation and prediction mode sharing to accelerate x265 encodings. Their approach substantially reduces time but remains tied to specific encoder implementations and presets. Menon~\etal~\cite{emes_ref} introduced heuristics-based schemes optimized for compression efficiency, encoding time savings, or a tradeoff between the two for HEVC-based streaming. Guo~\etal~\cite{av1_bayesian_multirate} applied Bayesian inference to AV1-based multi-rate encoding, using posterior probability distributions to infer block structures. Zaccarin~\etal~\cite{Zaccarin_ref} presented a DCT-domain multi-rate encoder that performs motion estimation only once for a reference stream, thereby reducing computational redundancy. However, adaptive drift error introduced to lower complexity can cause minor quality degradation, making it less ideal for highly dynamic content. Liu~\etal~\cite{liu_ref} proposed a VVenC-based fast multi-rate encoding approach that employs the partitioning structure from a low-bitrate reference encoding to accelerate the encoding of dependent representations. However, this work does not fully explore the potential of CU depth sharing across representations.

\textit{Key contributions:} 
We propose CU partitioning strategies, including single-bound, double-bound, and force partitioning, that effectively reuse CU depth information to minimize computational redundancy in multi-rate VVC encoding. To support these strategies, we develop a metadata-driven framework that enables efficient reuse of CU partitioning information across QPs, significantly reducing redundant operations. We evaluate the proposed methods across multiple VVenC presets. Furthermore, we assess the performance of methods through Pareto-front (PF) analysis, showcasing their ability to balance encoding efficiency and compression quality, thereby validating their scalability and practical applicability for adaptive streaming.

\section{Coding Unit Partitioning in VVenC}
VVC significantly enhances the flexibility and efficiency of CU partitioning compared to its predecessor, HEVC. VVC supports a wide range of CU sizes, from a maximum of $128\times 128$ to a minimum of $4\times 4$ for luma samples. This broad range enables better adaptation to varying video content characteristics, such as textures and motion complexity. To achieve this, VVC introduces six partitioning modes: Non-Split (NS), QuadTree (QT), Horizontal Binary Tree (HBT), Vertical Binary Tree (VBT), Horizontal Ternary Tree (HTT), and Vertical Ternary Tree (VTT)~\cite{vvc_block_partition_ref}. Each mode serves a distinct purpose: NS encodes the CU as a single block, while tree-based splits progressively divide the CU to capture finer details. Moreover, VVC includes advanced partitioning tools like Intra-Sub-Partitioning (ISP) and Geometric Partitioning (GEO)~\cite{geo_mode_ref}. ISP divides the CU into smaller sub-partitions, providing finer granularity for intra-prediction, which is particularly beneficial for capturing detailed textures. Conversely, GEO enables dynamic splitting based on geometric patterns, offering improved motion modeling in complex scenes. These features, when combined, provide greater precision and adaptability, making VVC highly effective at balancing compression efficiency and visual quality. This flexibility in CU partitioning also increases the computational complexity of the encoding process. 

\begin{figure}[t]
    \centering
\begin{subfigure}{0.480\columnwidth}
    \centering
    \includegraphics[width=\textwidth]{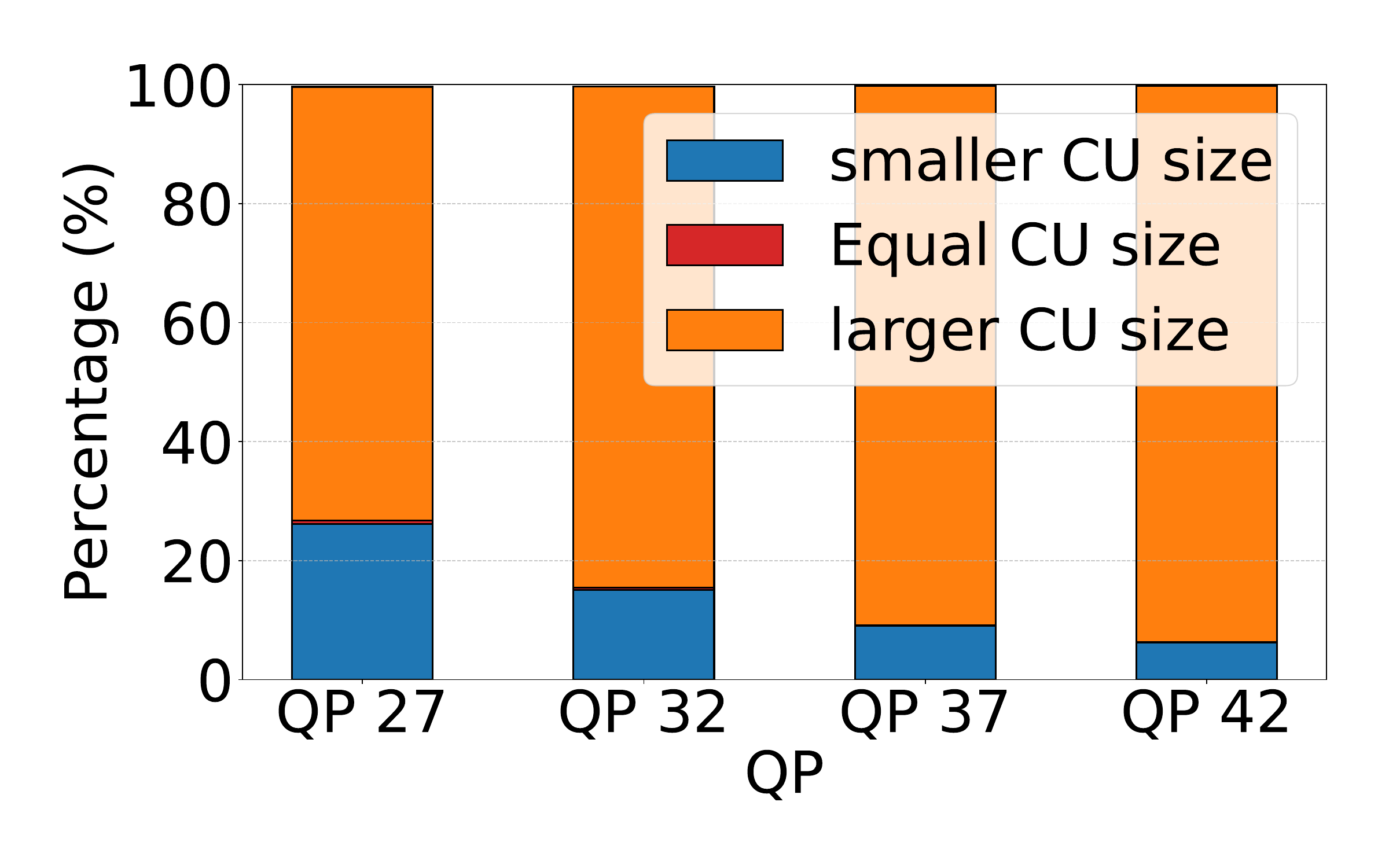}   
    \caption{\emph{BQTerrace}- \textit{medium}}
\end{subfigure}
\begin{subfigure}{0.480\columnwidth}
    \centering
    \includegraphics[width=\textwidth]{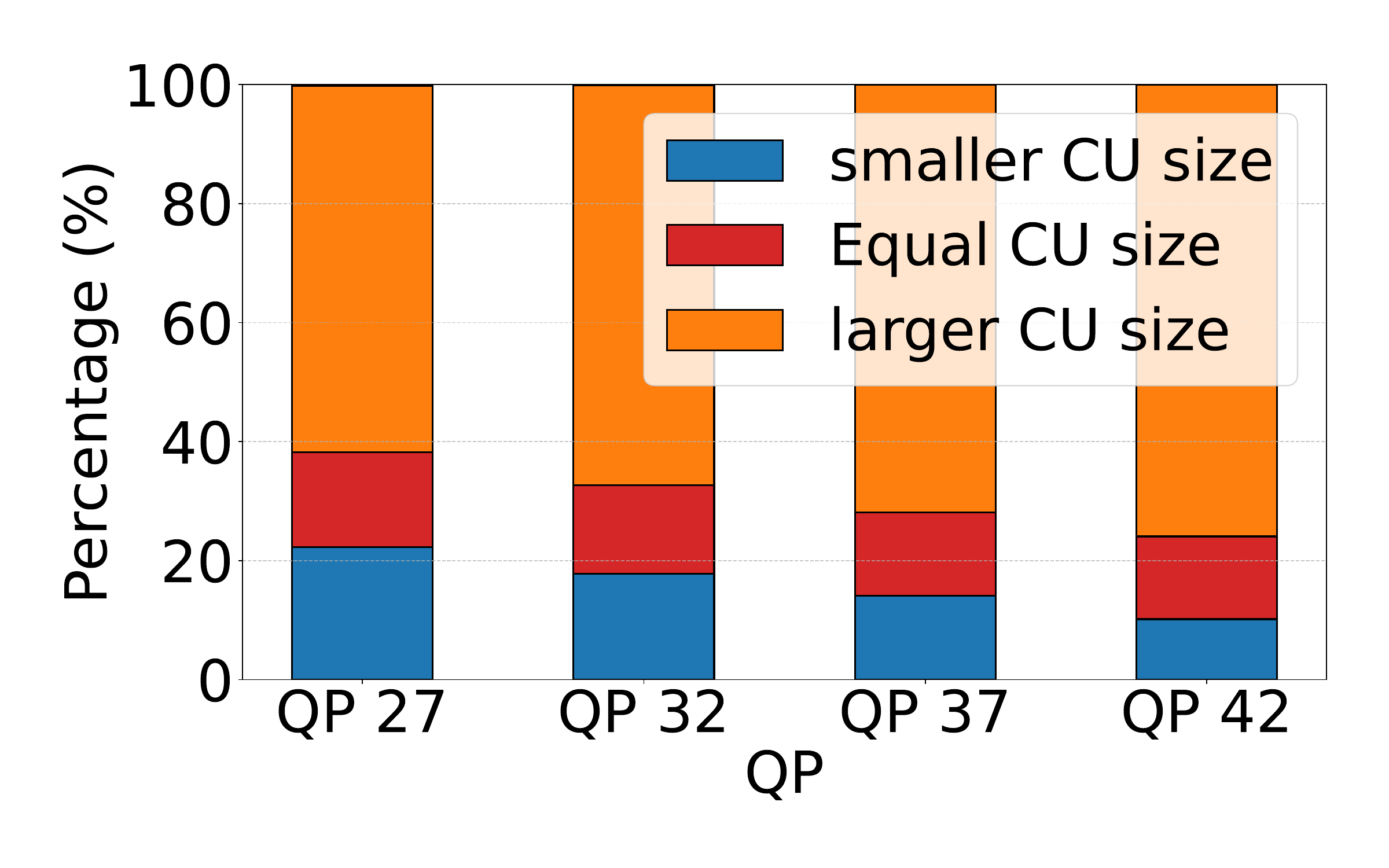}   
    \caption{\emph{CampfireParty}- \textit{medium}}
\end{subfigure}
\begin{subfigure}{0.480\columnwidth}
    \centering
    \includegraphics[width=\textwidth]{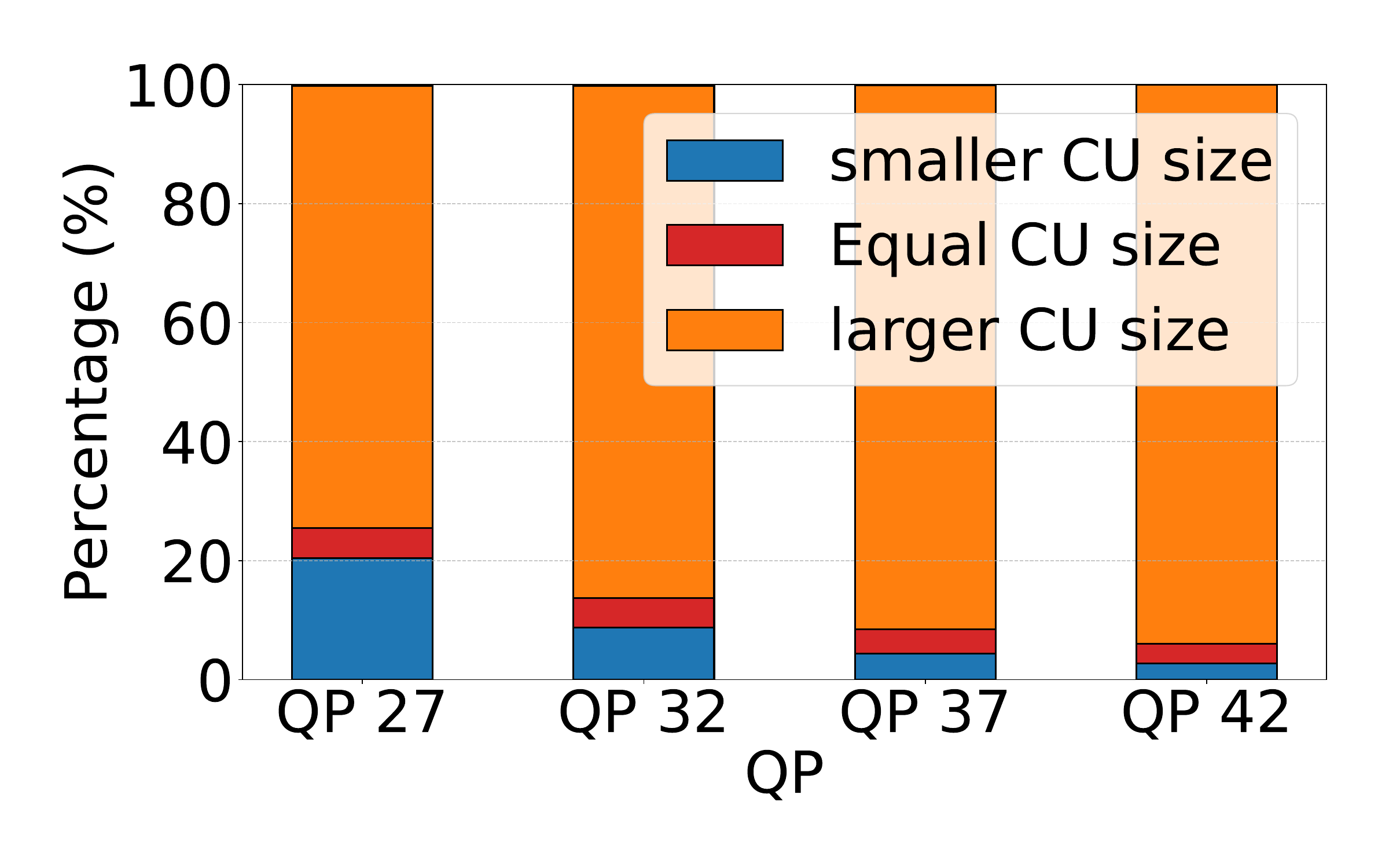}    
    \caption{\emph{BQTerrace}- \textit{faster}}
\end{subfigure}
\begin{subfigure}{0.480\columnwidth}
    \centering
    \includegraphics[width=\textwidth]{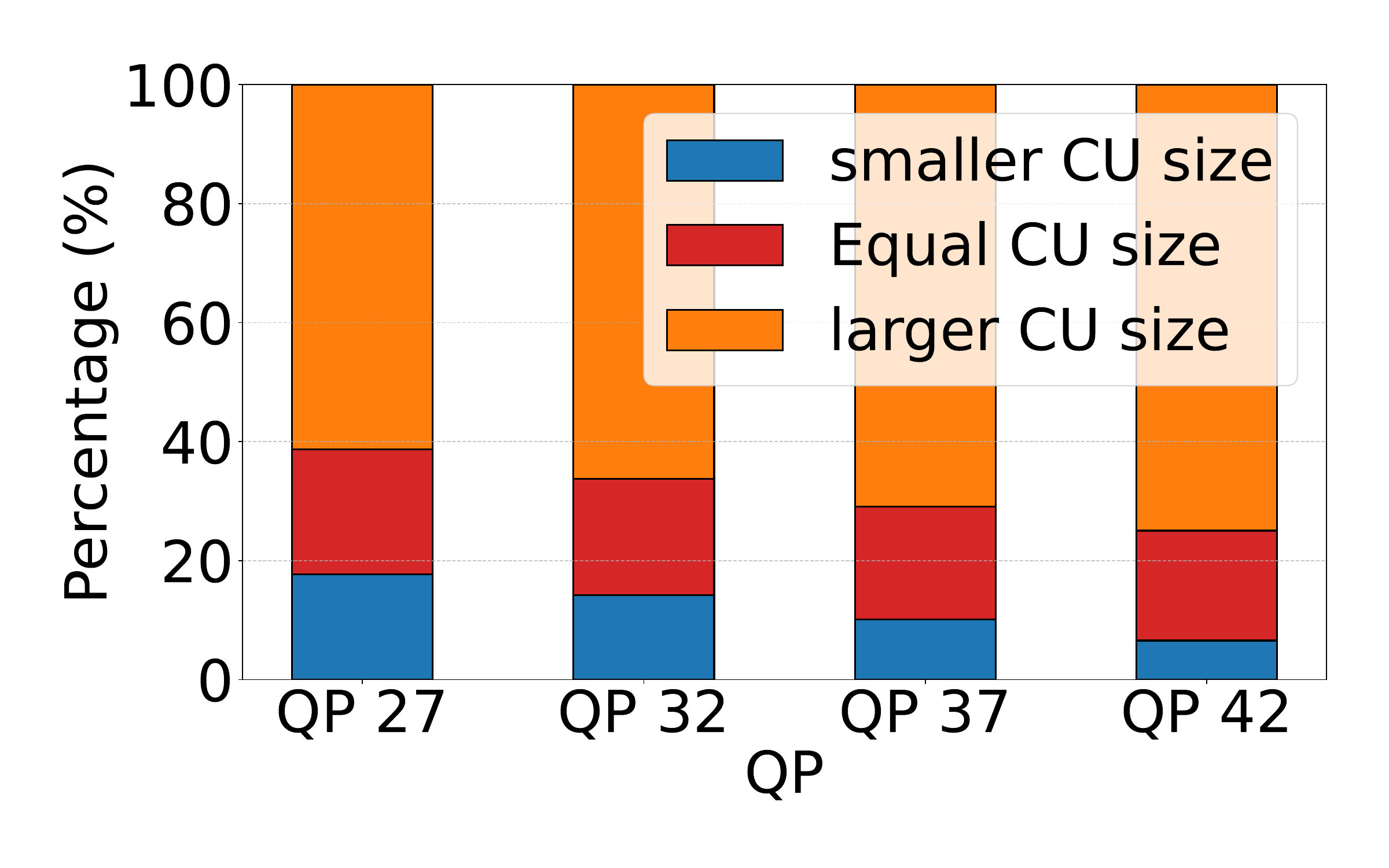}    
    \caption{\emph{CampfireParty}- \textit{faster}}
\end{subfigure}
\caption{CU depth statistics across various QP encodings compared to the QP22 encoding.}
\vspace{-0.99em}
\label{fig:depth_stats}
\end{figure}

VVenC adapts CU partitioning decisions to the complexity of video content, the target QP, and the encoding preset. Higher QPs generally produce shallower partitions due to increased quantization, while lower QPs result in deeper partitions to capture finer details. To study the CU depth correlation, we analyzed the partitioning patterns of encoded videos across a range of QPs (\eg 22, 27, 32, 37, and 42) and two presets~\cite{vvenc_preset_ref}, \emph{medium} and \emph{faster}, as shown in Fig.~\ref{fig:depth_stats}. For the \emph{medium} preset (Figs.~\ref{fig:depth_stats}a and \ref{fig:depth_stats}b), CU depths showed a strong correlation between neighboring QPs (\eg QP27 and QP32), with a high proportion of CU blocks retaining similar partitioning depths. However, as the QP difference increased (\eg QP22 vs. QP42), the correlation weakened, particularly for high QP encodes, which exhibited a noticeable reduction in CU partitioning depth. 
For the \emph{faster} preset (Figs.~\ref{fig:depth_stats}c and \ref{fig:depth_stats}d), a similar trend was observed, though with a greater tendency toward coarser partitions across all QPs. This behavior is consistent with the \emph{faster} preset’s focus on computational efficiency, leading to higher CU depth agreement for high QPs (\eg QP37 and QP42), but at the cost of finer-grained partitioning for low QPs (\eg QP22). High-bitrate encodes (lower QPs) consistently produced deeper partitions in both presets. In comparison, low-bitrate encodes (higher QPs) exhibited coarser partitions, with a split probability of approximately \SI{20}{\percent}--\SI{50}{\percent} for CUs at lower depths. These findings provide a quantitative basis for reusing CU depth information across QPs, supporting the proposed multi-rate encoding approaches and demonstrating trade-offs between partitioning granularity and computational efficiency across different encoding presets.

\section{Multi-rate encoding approaches}
\label{sec:proposed}

\subsection{Single-bound approaches}
\paragraph*{Top-Down Partitioning (\texttt{TDP})}
In this method, the CU depth obtained from the highest bitrate encode (\eg QP22) is treated as an \emph{upper bound} for all subsequent lower bitrate encodes. This ensures that the partitioning decisions made in the high-quality encode are reused, bypassing unnecessary finer-grained partitioning in lower-quality encodes.

Let:
\begin{itemize}
    \item $d(x, y)$: CU depth at position $(x, y)$ in the current encode.
    \item $d_{\text{HQ}}(x, y)$: CU depth at position $(x, y)$ in the highest bitrate encode (HQ).
    \item $\mathcal{D}_{\text{CTU}}$: Set of CU depths within a Coding Tree Unit (CTU).
\end{itemize}

The key constraint for this method is defined as:
\begin{equation}
    d(x, y) \leq d_{\text{HQ}}(x, y),
    \label{eq:upperbound}
\end{equation}
where $(x, y)$ represents the position of the CU within the CTU. This constraint ensures the current encode does not use finer splits than those in the highest bitrate encode.

\paragraph*{Bottom-Up Partitioning (\texttt{BUP})}
In this method, the CU depth from the lowest bitrate encode (\eg QP42) is processed first and treated as a \textit{reference} for all subsequent higher bitrate encodes. The CU depth from the reference encode serves as a \textit{lower bound}.

Let $d_{\text{LQ}}(x, y)$ be CU depth at position $(x, y)$ in the lowest bitrate encode (\text{LQ}). 
The key constraint for this method is defined as:
\begin{equation}
d(x, y) \geq d_{\text{LQ}}(x, y).
    \label{eq:lowerbound}
\end{equation}
This ensures that higher bitrate encodes cannot use shallower partitions than those in the lowest bitrate encode.

\subsection{Double-bound approaches}
\paragraph*{Bidirectional Constrained Partitioning (\texttt{BCP})}
In this method, the highest bitrate encode is processed first, providing the \textit{upper bound} for CU depth, as depicted in Eq.~\ref{eq:upperbound}. The lowest bitrate encode is processed next, using information from the highest bitrate encode, and its CU depth serves as the \textit{lower bound}, as shown in Eq.~\ref{eq:lowerbound}. 
For all intermediate bitrates, CU partitioning is constrained to lie \textit{within the bounds} of the highest and lowest bitrate encodes:
\begin{equation}
        d_{\text{LQ}}(x, y) \leq d(x, y) \leq d_{\text{HQ}}(x, y).
        \label{eq:db}
\end{equation}
This method combines the strengths of both \texttt{TDP} and \texttt{BUP}. The highest bitrate encode provides detailed partitioning, while the lowest bitrate encode prevents excessive splitting. Intermediate bitrates adapt within these bounds, ensuring a balance between quality and computational efficiency.

\paragraph*{Adaptive Hierarchical Partitioning (\texttt{AHP})}
In this method, the lowest bitrate encode is processed first, providing the \textit{lower bound} for CU depth as in Eq.~\ref{eq:lowerbound}. The highest bitrate encode is processed next, using information from the lowest bitrate encode, and its CU depth serves as the \textit{upper bound}, as shown in Eq.~\ref{eq:upperbound}. For all intermediate bitrates, CU partitioning is constrained to lie \textit{within the bounds} of the lowest and highest bitrate encodes as depicted in Eq.~\ref{eq:db}. Starting with the lowest bitrate ensures that coarse partitioning is established first. The highest bitrate encode then refines the partitions, adding detail only where necessary. Intermediate bitrates adapt within these bounds, gradually transitioning from coarse to fine partitioning.

\subsection{Force CU partitioning}

\paragraph*{Fixed Top-Down Reuse (\texttt{FTDR})}
The \texttt{FTDR} method enforces that CU partitioning decisions from the highest bitrate encode (\eg QP22) serve as both the lower and upper bound for all subsequent lower-bitrate encodes (QP27, QP32, \etc). This means that every CU in the dependent encodes must have the same depth as in the highest bitrate encode--neither finer nor coarser partitions are allowed. 
The \texttt{FTDR} constraint is defined as:
\begin{equation}
        d(x, y) = d_{\text{HQ}}(x, y).
\end{equation}
\texttt{FTDR} enforces the exact CU depth from the highest bitrate encode, restricting partitioning flexibility at lower bitrates. This leads to substantial encoding time savings but risks bitrate inefficiency due to the lack of rate-distortion adaptation.
 
\paragraph*{Fixed Bottom-Up Reuse (\texttt{FBUR})}
The \texttt{FBUR} method enforces that CU partitioning decisions from the lowest bitrate encode (\eg QP42) serve as the lower and upper bound for all subsequent higher-bitrate encodes.
The \texttt{FBUR} constraint is defined as:
\begin{equation}
        d(x, y) = d_{\text{LQ}}(x, y).
\end{equation}
\texttt{FBUR} eliminates the need for partitioning decisions at higher bitrates, substantially reducing encoding time. However, it can significantly degrade quality since high-bitrate encodes cannot refine partitions to capture finer details. This may lead to a higher BD-rate~\cite{DCC_BJDelta} compared to \texttt{FTDR}.

\section{Experimental Setup}
We conduct all experiments on a dual-processor server featuring Intel Xeon Gold 5218R processors (80 cores, operating at 2.10\,GHz), where each encoding instance is run on a single thread. We use random access configuration~\cite{jvet_ctc}, with an intra period of \SI{1}{\second}. To validate all encoding methods considered in this paper, we use the sequences from the classes A1, A2, and B of the JVET CTC~\cite{jvet_ctc}.

\begin{figure}[t]
\begin{center}
\includegraphics[width=0.9\columnwidth, trim=20 20 20 20, clip]{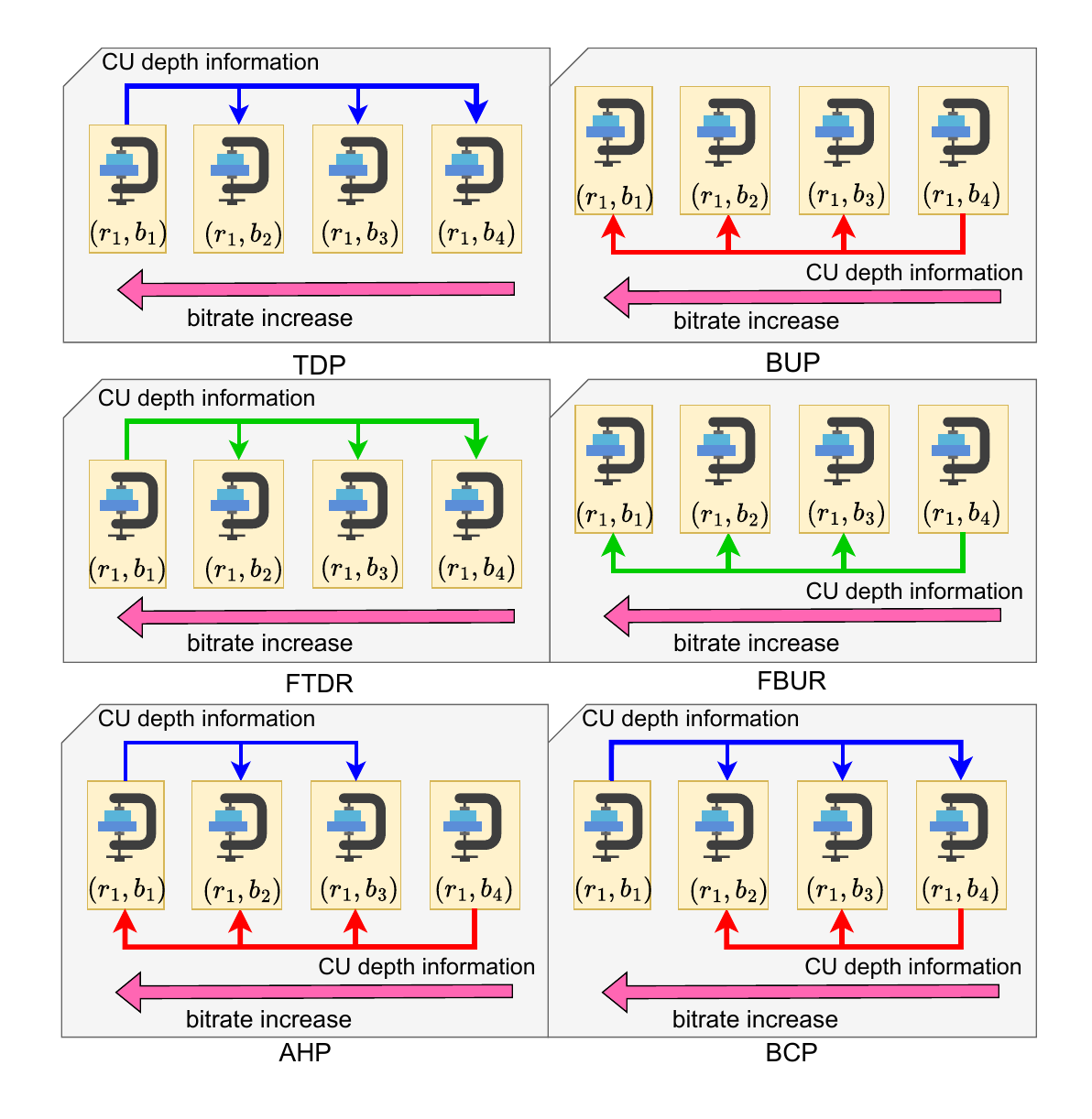}
\end{center}
\vspace{-0.9em}
\caption{CU depth analysis sharing methods for multi-rate encoding considered in this paper. The blue line denotes using the CU depth from the reference as a higher bound, the red line denotes using the CU depth from the reference as a lower bound, and the green line denotes forcing the same CU depth.}
\vspace{-0.9em}
\label{fig:arch}
\end{figure}

\paragraph*{Implementation} The multi-rate encoding framework, which reuses CU depth information across bitrate representations to minimize redundant computational effort, is implemented in VVenC\footnote{\url{https://github.com/vigneshvijay03/vvenc_multirate}}. Fig.~\ref{fig:arch} outlines the system's architecture, which integrates standalone VVenC encoders with mechanisms to share CU depth information across encodings. The CU depth information is stored in binary format to minimize storage overhead. Each CTU's partitioning data is encoded in a tree-like structure, where each node represents a CU and its associated depth. The metadata is indexed by CTU position $(x, y)$ to enable efficient retrieval during dependent encodings. Within the partitioning process, we have implemented mechanisms to incorporate constraints derived from metadata.

\paragraph*{Performance metrics} 
We compare the overall quality in PSNR~\cite{psnr_ref} and XPSNR~\cite{wien_xpsnr_vs_vmaf, xpsnr_ref} and the achieved bitrate for every target QP of each test sequence. Bjøntegaard Delta rate \BDRP~and \BDRX~measure the average bitrate overhead of the bitstreams compared with the \emph{Default} encoding at equivalent \mbox{PSNR} and \mbox{XPSNR}, respectively~\cite{DCC_BJDelta}.

To assess the computational efficiency of different multi-rate encoding methods, we evaluate both the serial ($T_{\text{S}}$) and parallel ($T_{\text{P}}$) encoding times. $T_{\text{S}}$ refers to the total cumulative encoding time required when each encoding process is executed sequentially, one after another. It is computed as:
\begin{equation}
    T_{\text{S}} = \sum_{i=1}^{N} \tau_i
\end{equation}
where $\tau_i$ is the encoding time for the $i^{th}$  bitrate representation, and $N$ is the total number of bitrates in the encoding ladder. This metric provides insight into the overall computational cost of encoding in a non-parallelized setting. $T_{\text{P}}$ reflects the encoding time in an ideal parallel execution scenario, where different bitrate encodings are processed simultaneously on separate computing resources. It is defined as:
\begin{equation}
    T_{\text{P}} = \max_{i=1}^{N} \tau_i
\end{equation}
This metric represents the maximum encoding time among all bitrates, assuming complete parallelism with sufficient hardware resources. We determine the average relative differences in serial encoding time (\delTS) and parallel encoding time (\delTP) relative to the \emph{Default} encoding.

\begin{figure}[t]
\centering
\begin{subfigure}{0.49\columnwidth}
    \centering
    \includegraphics[width=\textwidth]{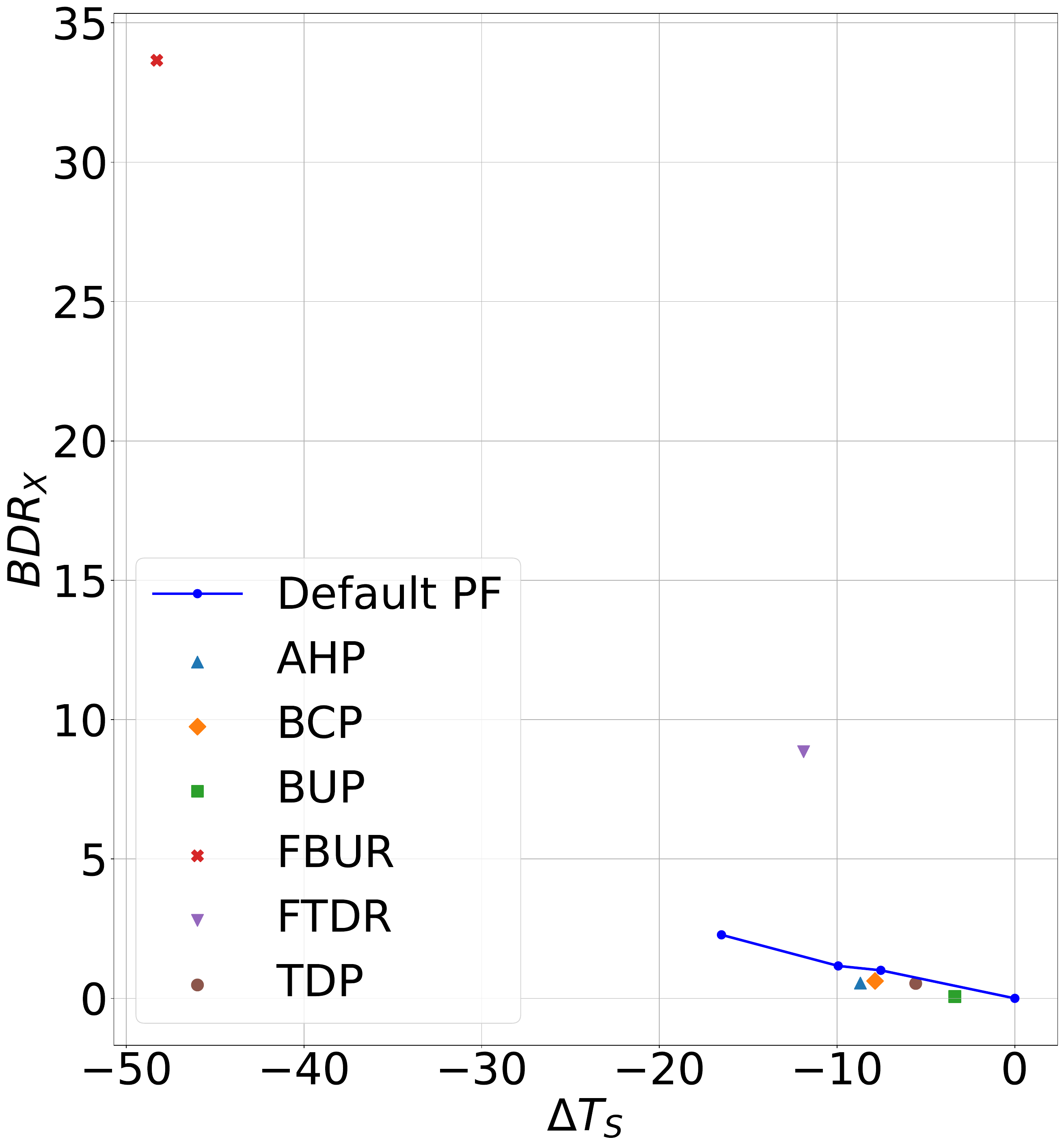} 
    \caption{$BDR_X$-$\Delta T_S$ \textit{medium}}
\end{subfigure}
\begin{subfigure}{0.49\columnwidth}
    \centering
    \includegraphics[width=\textwidth]{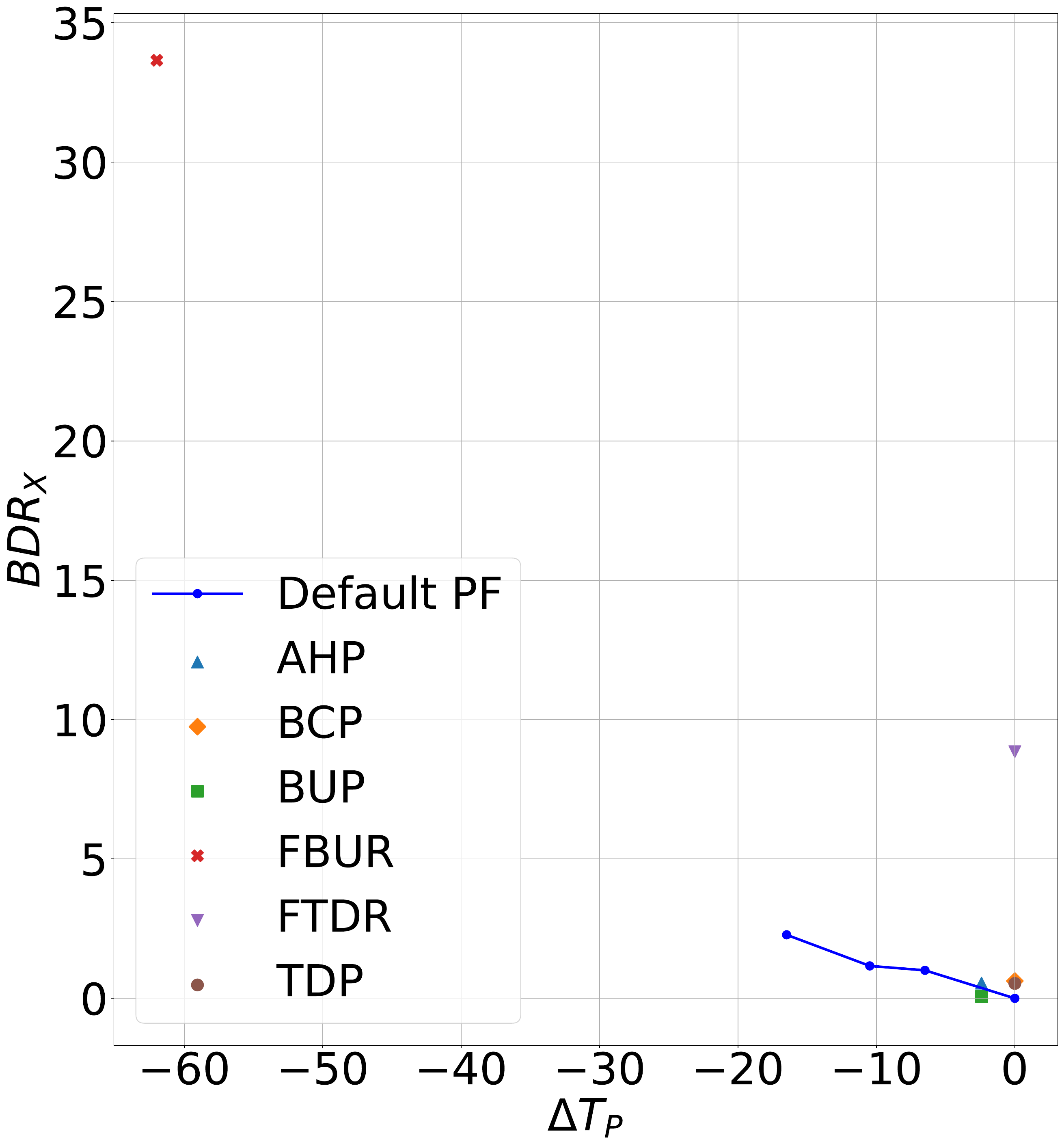} 
    \caption{$BDR_X$-$\Delta T_P$ \textit{medium}}
\end{subfigure}
\caption{$BDR_X$-$\Delta T$ results for the multi-rate encoding methods.}
\vspace{-0.99em}
\label{fig:pareto_res_ctc}
\end{figure}

\section{Experimental results}
\label{sec:results}
Fig.~\ref{fig:pareto_res_ctc} shows the relationship between bitrate savings (\BDRX) and encoding time savings ($\Delta T$) for different approaches and presets. The Default PF points for the \emph{medium} preset are evaluated using the configurations:
\begin{enumerate}
    \item \texttt{maxMTTDepthI=1}, which limits the maximum depth of MTT splits in intra-coded pictures to 1.
    \item \texttt{MIP=0, GEO-0, ISP=0}, which disables matrix-based intra prediction, geometric partitioning mode, and intra sub-partitions.
    \item \texttt{MIP=0, GEO=0, ISP=0, maxMTTDepthI=1}, which combines (1) and (2).
\end{enumerate}
It represents the baseline trade-off between bitrate efficiency and encoding complexity. An approach is deemed favorable if its points lie below and to the left of the PF, indicating improved encoding efficiency and reduced bitrate overhead.

\paragraph*{Bitrate savings vs. encoding time trade-off}
As observed in Table~\ref{tab:cqp_res}, \texttt{AHP} and \texttt{BCP} consistently exhibit the best trade-off, achieving significant encoding time reductions while maintaining low \BDRX increase. Meanwhile, \texttt{FTDR} and \texttt{FBUR} result in higher bitrate penalties, as they strictly enforce CU partitioning from reference encodes without adapting to finer rate-distortion optimizations. It is observed that the \delTP for \texttt{TDP}, \texttt{BCP}, and \texttt{FTDR} is $0$ since the highest bitrate encoding serves as the reference, and parallel encoding does not benefit from CU depth reuse in this case. However, \texttt{BUP}, \texttt{AHP}, and \texttt{FBUR} demonstrate time savings in \delTP due to their ability to accelerate high-bitrate encodings.

\paragraph*{Pareto-front analysis}
Fig.~\ref{fig:pareto_res_ctc} shows $BDR_X$--$\Delta T$ results for the multi-rate encoding methods using \textit{Default} as the reference point for JVET sequences and CTC configuration. Methods below the PF (\eg \texttt{AHP}, \texttt{BCP}) exhibit the best trade-off between encoding time savings and bitrate increase.  \texttt{BCP} and \texttt{AHP} remain the closest to the PF, meaning they provide the best trade-off between encoding time savings and an acceptable bitrate increase. \texttt{TDP} and \texttt{BUP} achieve moderate efficiency gains but do not match the optimal performance of \texttt{BCP} and \texttt{AHP}. \texttt{FBUR} and \texttt{FTDR}, while aggressively reducing encoding time, fall far short of the PF due to their high bitrate penalties.

\paragraph*{Practical recommendation} Based on our findings, we recommend the \texttt{AHP} or \texttt{BCP} methods for practical adaptive streaming use-cases, as they offer the best balance of encoding speed and minimal BD-rate loss. These methods can be safely adopted in VoD encoding pipelines without compromising quality and achieving substantial compute savings.

\section{Conclusions}
\label{sec:conclusion_future_dir}
This paper presented CU partitioning strategies to improve the efficiency of multi-rate encoding for adaptive video streaming. The proposed framework effectively reduces computational redundancy while maintaining high video quality by leveraging single- and double-bound approaches and force partitioning methods. Experimental results across multiple presets and sequences demonstrate significant encoding time serial and parallel encoding time reductions with savings of up to \SI{11.69}{\percent} and minimal bitrate overhead ($<$\SI{0.6}{\percent}). PF analysis highlights the superior performance of \texttt{BCP} and \texttt{AHP}, showcasing their ability to achieve a balanced trade-off between time savings and compression efficiency. Overall, our CU-guided multi-rate encoding framework provides a scalable path forward for next-generation video services that seek to balance quality and efficiency. By integrating our strategies into open-source encoders such as VVenC, the proposed methods can be used in cloud VoD encoding pipelines, where even a 10\% average time savings per title translates into significant reductions in compute costs and energy use.

Future work will extend these strategies to multi-resolution encoding, ensuring efficient CU depth reuse across both bitrates and resolutions. Additionally, we will investigate the computational redundancies regarding mode decisions and motion estimation. 
\begin{table}[t]
\caption{Average encoding performance compared to the \textit{Default} encoding for classes A1, A2 and B of JVET CTC.}
\centering
\resizebox{0.9\columnwidth}{!}{
\begin{tabular}{@{}l@{ }|@{ }c@{ }|@{ }c@{ }|@{ }c@{ }|@{ }c@{ }|@{ }c@{ }|@{ }c@{ }|@{ }c@{ }}
\specialrule{.12em}{.05em}{.05em}
\specialrule{.12em}{.05em}{.05em}
Preset & Method & \BDRP & \BDRX & \mbox{BD-PSNR} & \mbox{BD-XPSNR} & \delTS & \delTP \\
& & [\%] & [\%] & [dB] & [dB] & [\%] & [\%]  \\
\specialrule{.12em}{.05em}{.05em}
\specialrule{.12em}{.05em}{.05em}
\multirow{5}{*}{\emph{medium}}  & \texttt{TDP}	&	0.53	&	0.51	& -0.02 & -0.01 & 	-6.57	&	0\\
                        & \texttt{BUP}~\cite{liu_ref}	&	0.06	&	0.05	& 0 & 0 &	-3.37	&	-2.41\\
                        & \texttt{BCP}	&	0.62	&	0.59	& -0.02 & -0.01 & 	-9.87	&	0\\
                        & \texttt{AHP}	&	0.54	&	0.50	& -0.02 & -0.01 & 	-11.69	&	-2.41\\
                        & \texttt{FTDR}	&	8.85	&	8.65	& -0.20 & -0.07 &	-13.89	&	0\\
                        & \texttt{FBUR}	&	33.65	&	32.14	& -0.87 & -0.15 &	-50.28	&	-61.98\\
\hline
\multirow{5}{*}{\emph{fast}} & \texttt{TDP}	  &	  0.66	  &	  0.61	  &	 -0.02 & -0.01 &  -4.93	  &	  0\\
                      & \texttt{BUP}~\cite{liu_ref}	&	0.27	&	0.24	& -0.01 & -0.01 & 	-3.11	&	-2.21\\
                      & \texttt{BCP}	&	0.90	&	0.82	& -0.03 & -0.01 &	-7.93	&	0\\
                      & \texttt{AHP}	&	0.81	&	0.75	& -0.02 & -0.01 &	-7.38	&	-2.21\\
                      & \texttt{FTDR}	&	22.66	&	21.38	& -0.58 & -0.16 &	-11.38	&	0\\
                      & \texttt{FBUR}	&	32.50	&	30.89	& -0.86 & -0.12 & 	-38.45	&	-50.95\\
\hline
\multirow{5}{*}{\emph{faster}} & \texttt{TDP}	&	0.60	&	0.59	& -0.02 & -0.01 &	-3.70	&	0\\
                        & \texttt{BUP}~\cite{liu_ref}	&	0.25	&	0.23	& -0.01 & -0.12 & 	-0.11	&	-0.03\\
                        & \texttt{BCP}	&	0.85	&	0.81	& -0.03 & -0.01 &	-4.41	&	0\\
                        & \texttt{AHP}	&	0.76	&	0.71	& -0.03 & -0.01 &	-3.45	&	-0.03\\
                        & \texttt{FTDR}	&	22.45	&	21.39	& -0.57 & -0.18 &	-10.01	&	0\\
                        & \texttt{FBUR}	&	32.90	&	31.04	& -0.87 & -0.14 &	-37.97	&	-54.52\\
\specialrule{.12em}{.05em}{.05em}
\specialrule{.12em}{.05em}{.05em}
\end{tabular}}
\label{tab:cqp_res}
\end{table}

\balance
\bibliography{references.bib}{}
\bibliographystyle{IEEEtran}

\balance
\end{document}